\documentstyle [preprint,aps] {revtex}
\begin{document}
\draft
\title{ Kinetic Approach to Fractional Exclusion Statistics }
\author{ G. Kaniadakis
\thanks{e-mail: Kaniadakis@pol88a.polito.it},
 A. Lavagno and P. Quarati}
\address{ Dipartimento di Fisica - Politecnico di Torino -
Corso Duca degli Abruzzi 24, 10129 Torino, Italy \\
Istituto Nazionale di Fisica Nucleare, Sezioni di Cagliari
e di Torino\\
Istituto Nazionale di Fisica della Materia,
Unit\'a del Politecnico di Torino}
\date{\today}
\maketitle
\begin {abstract} {\bf Abstract:}
We show that the kinetic approach to statistical mechanics
permits an elegant and
efficient treatment of fractional exclusion statistics. By using the
exclusion-inclusion principle recently proposed
[Phys. Rev. E {\bf 49}, 5103 (1994)] as a
generalization of the Pauli exclusion principle, which is based
on a proper definition of the transition probability between
two states, we derive a variety of different
statistical distributions interpolating between bosons and
fermions. The Haldane exclusion principle and the Haldane-Wu
fractional exclusion statistics are obtained in a natural way as
particular cases. The thermodynamic properties of the statistical
systems obeying the generalized exclusion-inclusion principle
are discussed.
\end {abstract}

Following Haldane's formulation \cite{Hal} of a generalized
Pauli exclusion principle, many papers have been recently
devoted to the study of fractional exclusion statistics by
interpolation of bosonic and fermionic distributions \cite{Wu}.
There is an intrinsic connection between these fractional
statistics and the interpretation of the fractional quantum
Halleffect \cite{Lau} and anyonic physics \cite{Fra,Com}.
 Murthy and Shankar \cite{Murt1} generalized the Haldane
statistics to infinite dimensions and showed \cite{Murt2} that
the one dimensional bosons interacting through the two-body
inverse square potential $V_{ij}=g(g-1)(x_i-x_j)^{-2}$ of
the Calogero-Sutherland model obeys fractional exclusion
statistics in the sense of the Haldane interpretation.
  Using the Thomas-Fermi method, Sen and Bhaduri
\cite{Sen} considered the particle exclusion statistics for the
Calogero-Sutherland model in the presence of an external,
arbitrary one-body potential. Isakov
\cite{Isak} extended the Calogero-Sutherland model in the case
of particles interacting through a generic two-body potential
$V_{ij}$ and derived the related exclusion statistics. Rajagopal
\cite{Raja} obtained the von Neumann entropy per
state of the Haldane exclusion statistics. Nayak and Wilczek
\cite{Nayak} studied the low-temperature properties, the
fluctuations and the duality of exclusions
statistics. Murthy and Shankar \cite{Murt1} also computed, for
quasi-particles, in the Luttinger model, the exclusion statistics
parameter $g$ related to the exchange statistical parameter
$\alpha$ of the quantum phase $e^{i\pi\alpha}$, and  showed
\cite{Murt2} that the parameter $g$ can be completely
determined by the second virial coefficient in the high
temperature limit.

Haldane defined the statistics $g$ of a particle by
\begin{equation}
g=-\frac{{\rm d}_{_{N+\Delta N}}-{\rm d}_{_N} }{\Delta N}  \ \ ,
\end{equation}
where $d_{_N}$ is the single-particle Hilbert-space  dimension,
when the
coordinates of $N-1$ particles are kept fixed.
Thus the dimension of the Hilbert space for the single particle
states is a finite and
extensive quantity that depends on the number of particles in
the system. One can choose this dimension as
\begin{equation}
{\rm d}_{_N}=K-g (N-1) \ \ ,
\end{equation}
and the statistical weight or degeneracy factor is
\begin{equation}
W=\frac{({\rm d}_{_N}+N-1) ! }{ N ! \, ({\rm d}_{_N}-1) ! } \ \ .
\end{equation}
This equation is a simple interpolation between the number of
ways of placing $N$ identical bosons or fermions in $K$
single-particle independent states confined to a
finite region of matter. The expression of the mean occupation
number  $n=N/K$  has been obtained  in an
implicit form by Wu \cite{Wu}.
\begin{equation}
n=\frac{1}{w(E)+g} \ \ ,
\end{equation}
with $w$ satisfying the relation
\begin{equation}
w(E)^g \, [1+w(E)]^{1-g}=e^{\beta (E-\mu)} \ \ .
\end{equation}
In the special cases $g=0$ and $g=1$,  Eq.(4) yields, respectively,
the Bose-Einstein (BE) and Fermi-Dirac (FD) distributions.

Ilinski and Gunn \cite{Ilin} criticize the identity of relations (1)
and (3) of Haldane and, using (2), derive a different $W$
constructing a statistical mechanics which is
not in complete agreement with Wu statistical mechanics
\cite{Wu}.

Acharya and Swamy \cite{Achar} and Polychronakos \cite{Poly}
have studied, in addition to the Haldane exclusion statistics, new
fractional statistics with appealing
features as positive probabilities and analytical expressions for
all termodynamic quantities. We wish to recall the first work on
intermediate statistics published by
Gentile at the beginning of 1940 \cite{Gentile}.

In a previous work \cite{Ka} we have considered the kinetics of
particles in a phase space of arbitrary dimensions $D$, obeying
an exclusion-inclusion principle.
We obtained the statistical distributions of the particle system
as the stationary state of a non-linear kinetic equation.
A crucial point of this formalism is the definition of the
transition probability which can be written in various forms,
containing the effects of the inclusion-exclusion
principle through the distribution function $n=n(t,{\bf v})$,
which is the mean occupation number of the state ${\bf v}$.

Setting $\pi(t,{\bf v} \rightarrow {\bf u})$ the transition
probability from the state ${\bf v}$ to the state ${\bf u}$,
the evolution equation of the distribution function
$n(t,{\bf v})$ can be written as
\begin{equation}
\frac{\partial n(t,{\bf v})}{\partial t}=\int [\pi(t,{\bf u}
\rightarrow {\bf v})-
\pi(t,{\bf v} \rightarrow {\bf u})] \, d^{_{^D}} u \ \ .
\end{equation}
In ref.\cite{Ka} we postulated the following expression of the
transition probability :
\begin{equation}
\pi(t,{\bf v} \rightarrow {\bf u})=r(t,{\bf v},{\bf v}-{\bf u}) \,
\varphi[n(t,{\bf v})] \,  \psi[n(t,{\bf u})] \ \ ,
\end{equation}
where $r(t,{\bf v},{\bf v}-{\bf u})$ is the transition rate,
 $\varphi[n(t,{\bf v})]$ is a function depending on the
 occupational distribution at the initial state ${\bf v}$
and $\psi[n(t,{\bf u})]$ depends on the arrival state.
The function $\varphi(n)$ must obey the condition
$\varphi(0)=0$ because the
transition probability is equal to zero if the initial state is empty.
Furthermore, the function $\psi(n)$ must obey the condition
$\psi(0)=1$ because, if the arrival state
is empty, the transition probability is not modified. If we choose
$\varphi(n)=n$
and $\psi(n)=1$ we obtain the standard linear kinetics. The
function $\psi(n)$ defines implicitely the inclusion-exclusion
principle enhancing or inhibiting the transition probability.

In this letter we use the kinetic approach outlined above to
generate new  fractional exclusion statistics. The approach is
appropriate both for non-interacting and interacting particles.
 In the following, we derive general expressions linking the
main statistical and thermodynamic quantities to the transition
probabilities of the kinetic theory. We obtain implicitely a
fractional statistics which interpolates between BE and FD
distributions in a single-particle Hilbert space whose dimension
is an arbitrary function of $n$. We show that the Haldane-Wu
statistics can be derived from this one by considering Brownian
particles and by demanding that the Hilbert space
dimension be a linear function of $n$. Another particular case
within the family of statistics here introduced yields a statistics
interpolating among BE, FD and Maxwell-Boltzmann (MB)
distributions.

Let us  consider Eqs. (6) and (7) and limit the discussion to the
one-dimensional velocity space (the extension to a
$D$-dimensional phase space is straightforward).
We limit ourselves to first neighbors interactions, this being
equivalent to truncate the Moyal expansion given by Eq.(7)
of Ref. \cite{Ka} at the second order and we
obtain the following, generalized, non-linear
Fokker-Planck equation
\begin{eqnarray}
\frac{\partial n(t,v)}{\partial t}=\frac{\partial}{\partial v}
\Bigg [\left (J(t,v)+\frac{\partial D(t,v)}{\partial v}\right )
\varphi(n) \psi(n) \ \ \
\nonumber \\+ D(t,v) \left (\psi(n) \frac{\partial
\varphi(n)}{\partial n}-\varphi(n)
\frac{\partial \psi(n)}{\partial n}\right ) \frac{\partial
n(t,v)}{\partial v} \Bigg ]  \ \ .
\end{eqnarray}
$J(t,v)$ and $D(t,v)$ are the drift and diffusion coefficients,
respectively, and are given by the first and the second order
moments of the transition rate \cite{Ka}.
Eq.(8) is a continuity equation for the distribution function
$n=n(t,v)$
\begin{equation}
\frac{\partial n(t,v)}{\partial t}+\frac{\partial j(t,v,n)}
{\partial v}=0 \ \ ,
\end{equation}
where the particle current $j=j(t,v,n)$ is given by
\begin{eqnarray}
j=-D \, \frac{\varphi(n) \psi (n)}{\overline{(\Delta n)^2}}
\Bigg [\frac{\partial \epsilon}{\partial v} \,
 \overline{(\Delta n)^2}
+ \frac{\partial n}{\partial v} \Bigg ] \ \ .
\end{eqnarray}
We have defined
\begin{equation}
\overline{(\Delta n)^2}=\left \{ \frac{\partial}{\partial n} \log
\left [ \frac{\varphi(n)}{\psi(n)}  \right] \right \}^{-1} \ \ ,
\end{equation}
and
\begin{equation}
\frac{\partial \epsilon}{\partial v}=\frac{1}{D(t,v)}\left (J(t,v)
+\frac{\partial D(t,v)}{\partial v}\right )  \ \ .
\end{equation}
The function $\epsilon =\epsilon (v)$ is an adimensional single
particle energy defined up to an additive, arbitrary constant.
 This energy is appropriate both for non-interacting and
interacting particles. Nayak and Wilczek \cite{Nayak} have
examined the particular case $\epsilon (v) \propto v^l$ with
$l$ integer. The case $l=2$ corresponds to Brownian particles,
 the drift and diffusion coefficients being
given by
\begin{equation}
J=\gamma v, \, \, \, \, \, \,\, \, \, \, \, \,
D=\frac{\gamma}{\beta m} \ \ .
\end{equation}
The quantity $\gamma$ is a dimensional constant and
$\epsilon$ is given by: $\epsilon=\beta (E-\mu)$ with
$E=\frac{1}{2} m v^2$ being the kinetic energy.
$\beta=1/T$ ($k_{_B}=1$) is the inverse of the temperature and
$\mu$ is the chemical potential.

In stationary conditions, the particle current vanishes:
$j(\infty,v,n)=0$. Eq.(10) becomes a omogeneous first order
differential equation,
which can be integrated easily:
\begin{equation}
\frac{\psi(n)}{\varphi(n)}=\exp (\epsilon) \ \ .
\end{equation}
When the functions $\varphi(n)$ and $\psi(n)$ are fixed, Eq.(14)
 gives the statistical distribution $n=n(\epsilon)$ of the system.

We stress that, if we use as variable the single particle energy
$\epsilon$, it is easy
to verify that the above-defined function
$\overline{(\Delta n)^2}=<n^2>-<n>^2$,
is equal to the second order density fluctuation \cite{Landau}
\begin{equation}
\overline{(\Delta n)^2}=- \frac{\partial n}{\partial\epsilon} \ \ .
\end{equation}
This relation is very important because it reveals that the mean
fluctuation of the occupation number is the crucial quantity that
determines the equilibrium distribution.

The statistical distribution $n(\epsilon)$ can be obtained by the
maximum entropy principle, fixing the total number of
particles and energy of the system and using the standard
Lagrange multiplier method. Setting ${\cal S}(n)=S(N)/K$
the thermodynamic limit of the entropy per state, we have
\begin{equation}
\frac{\partial}{\partial n}\left [{\cal S}(n)-\epsilon \,
 n \right ]=0\ \ ,
\end{equation}

{}From Eqs. (14) and (16) it is possible to obtain the entropy in
terms of the functions $\varphi(n)$ and $\psi(n)$ as
\begin{equation}
\frac{\psi(n)}{\varphi(n)}=\exp\left [\frac{\partial {\cal S}(n)}
{\partial n} \right ] \ \ ,
\end{equation}
or in terms of $\overline{(\Delta n)^2}$
\begin{equation}
\overline{(\Delta n)^2}=-\left [ \frac{\partial ^2 {\cal S}}
{\partial n^2} \right ]^{-1} \ \ .
\end{equation}
By using the Boltzmann principle with the identification of the
entropy ${\cal S}=\log {\cal W}$, it is possible to find the
relation between ${\cal W}$ and the functions $\varphi(n)$ and
$\psi(n)$ in the following expression
\begin{equation}
\frac{\psi(n)}{\varphi(n)}=\exp\left [ \frac{1}{{\cal W}(n)}
\frac{\partial {\cal W}(n)}{\partial n} \right ] \ \ .
\end{equation}

It is well known that the partition function $\cal{Z}$ per state
\begin{equation}
{\cal Z}=e^{- \beta\Omega} \ \ ,
\end{equation}
is related to the mean density $n$ by means of
\begin{equation}
n=-\frac{\partial}{\partial \epsilon}  \log {\cal Z} \ \ .
\end{equation}
Taking into account Eqs. (11), (15), (20) and (21), we may
write the thermodynamic potential $\Omega$ as a function of
$\varphi(n)$ and $\psi(n)$
\begin{equation}
\beta\Omega=\int n \, \frac{\partial}{\partial n}
\log\frac{\psi(n)}{\varphi(n)} \, dn \ \ ,
\end{equation}
and the density fluctuation as
\begin{equation}
\overline{(\Delta n)^2}=-\, n \, \left [ \frac{\partial}{\partial n}
\beta \, \Omega  \right ]^{-1} \ \ .
\end{equation}

Now let us consider the case $\varphi(n)=n$;  for the function
$\psi(n)$ we choose the particular form
\begin{equation}
\psi=\displaystyle{d^{-\frac{\partial d}{\partial n}} \,
(n+d)^{1+\frac{\partial d}{\partial n}}}  \ \ .
\end{equation}
where the function $d=d(n)$ must satisfy the condition $d(0)=1$.
 In this case, Eq.(14) which defines the statistics, becomes:
\begin{equation}
d^{-\frac{\partial d}{\partial n}} \, (n+d)^{1+
\frac{\partial d}{\partial n}} = n \, e^{\epsilon} \ \ .
\end{equation}
 The function ${\cal W}$ becomes
\begin{equation}
{\cal W}=  \frac{(n+ d)^{n+d}}{n^n \,  d^d} \ \ ,
\end{equation}
and the entropy
\begin{equation}
{\cal S}=- n [ w \log w - (1+w) \log (1+w) ] \ \ ,
\end{equation}
where $w=d/n$, so we find the general form of the
von Neumann entropy per state
\cite{Raja}. The thermodynamic potential $\Omega$ becomes
\begin{equation}
\beta\Omega=\left (n \, \frac{\partial d}{\partial n}-d \right ) \,
\log \frac{n+d}{d} \ \ ,
\end{equation}
and the partition function ${\cal Z}$ can be rewritten as
\begin{equation}
{\cal Z}=\left (  \frac{n+d}{d} \right )^{d-n\, \frac{\partial d}
{\partial n}} \ \ ,
\end{equation}
while  the fluctuation becomes
\begin{equation}
\overline{(\Delta n)^2}=\frac{n \, d \, (n+d)}{\left (d-n \,
\displaystyle{\frac{\partial d}{\partial n}}\right )^2 -n \, d\,
(n+d) \, \displaystyle{ \frac{\partial ^2 d}{\partial n^2}} \,
\log\frac{n+d}{d}   } \ \ .
\end{equation}

The quantity $d$ is an arbitrary function and we can use it to
define a family of exclusion statistics. If we select a particular
form of $d$, we observe that this one can depend not only on
$n$ but also on a parameter $g$: $d=d(g,n)$. In this way
Eq.(25) defines, varying $g \in [0,1]$, a fractional statistics.
If we require that the statistics interpolates between BE and FD
distributions we must set the two conditions: $d(0,n)=1$ and
 $d(1,n)=1-n$. These two conditions imply that  the
single-particle Hilbert-space  dimension be that of the Bose
space when $g=0$ and that of the Fermi one when $g=1$.
Let us call $S(N)=K{\cal S}(n)$ and $W(N)$ the entropy and
statistical weight of the system of $N$ particles lying in $K$
states. From the Boltzmann principle
$S=\log W$ we obtain ${\cal W}=W^{1/K}$. It is easy to see
that ${\cal W}$, given by Eq.(26), is the thermodynamic limit of
the statistical weight given by Eq. (3), where, this time,
${\rm d}_N={\rm d}(g,N,K)$ is the single-particle Hilbert-
space  dimension, is an arbitrary function and admits as
thermodynamic limit
$d(g,n)={\rm d}(g,N,K)/K$. If we consider a statistics
interpolating between bosons and fermions, the conditions to
which  $d=d(g,n)$ must satisfy when
$g=0,1$ can be derived in the thermodynamic limit from the
two conditions ${\rm d}(0,N,K)=K$ and ${\rm d}(1,N,K)=K-N+1$.

The Haldane-Wu choice of ${\rm d}$ given by Eq.(2) implies
\begin{equation}
d=1-g \, n \ \ ,
\end{equation}
and requires that $\psi (n)$ be given by
\begin{equation}
\psi(n)=[1-g\, n]^g \, [1+(1-g) \, n]^{1-g} \ \ .
\end{equation}
 Equation (25) becomes
\begin{equation}
[1-g\, n]^g \, [1+(1-g) \, n]^{1-g}=n \, e^{\epsilon} \ \
\end{equation}
and we obtain, in the case of Brownian particles, the statistics
introduced by
Haldane \cite {Hal} and by Wu \cite{Wu}.
The partition function ${\cal Z}$ and the density fluctuation are
given by
\begin{equation}
{\cal Z}=\frac{1+(1-g) n}{1-gn} \ \ .
\end{equation}
\begin{equation}
\overline{(\Delta n)^2}= n \, (1-g n) \, [1+(1-g) n] \ \ .
\end{equation}

As a second example of fractional statistics we consider the
statistics defined by:
\begin{equation}
n=\frac{1}{\exp(\epsilon)-\kappa} \ \ ,
\end{equation}
studied extensively in ref. \cite{Achar,Poly,Ka}. For
$\kappa=-1$, $0$ and $1$ one obtains the FD, MB and BE
statistical distributions, respectively. In ref.
\cite{Achar} it is shown that the parameter $\kappa$ is a
function of the exchange statistical parameter $\alpha$
appearing in the quantum phase $e^{i \pi\alpha}$.
This statistics is generated from
\begin{equation}
\psi(n)=1+\kappa n \ \ .
\end{equation}
In this case the partition function can be written as
\begin{equation}
{\cal Z}=(1+\kappa n)^{1/\kappa} \ \ ,
\end{equation}
and it is easy to verify that the density fluctuation
$\overline{(\Delta n)^2}$ and the
entropy ${\cal S}$ are given by
\begin{equation}
\overline{(\Delta n)^2}= n (1+\kappa n) \ \ ,
\end{equation}
\begin{equation}
{\cal S}=\frac{1}{\kappa} (1+\kappa n ) \log (1+\kappa n)- n
 \log n \ \ ,
\end{equation}
while the statistical weight ${\cal W}$ becomes
\begin{equation}
{\cal W}=\frac{ (1+\kappa n)^{ (1+\kappa n)/\kappa} } { n^n} \ \ ,
\end{equation}
and can be obtained as the thermodynamic limit of the statistical weight:
\begin{eqnarray}
W &=& |\kappa|^N \, \frac{\left (
\displaystyle{\frac{K}{|\kappa|}}\right ) !}
{N !  \, \left ( \displaystyle{\frac{K}{|\kappa|}}-N \right ) ! }  \ \ ,
 \ \ \kappa <0 \ \ , \\
W &=& \frac{K^N}{N ! }  \ \ , \ \ \ \ \ \ \ \ \ \ \ \ \ \ \ \
\ \ \ \ \ \ \kappa =0 \ \ , \\
W &=& \kappa^N \, \frac{\left (\displaystyle{\frac{K}{\kappa}}+
N-1\right ) !}
{N !  \, \left ( \displaystyle{ \frac{K}{\kappa}}-1\right ) ! }  \ \ ,
 \ \ \kappa >0 \ \ .
\end{eqnarray}

In conclusion, we have shown that the effects of the
exclusion-inclusion principle can be taken into account in the
particle kinetics by means of a proper definition of
the transition probability. We have derived a non-linear
evolution equation which admits as steady solutions the
exclusion statistical distributions. We have linked the
main thermodynamic properties of the system with the function
$\psi (n)$ which generates the exclusion statistics. As first
 application of the theory we have considered a family of
exclusion statistics in which the single-particle Hilbert-space
dimension is an arbitrary function of the mean occupational
number and contains, as particular case, the Haldane-Wu
statistics. As second application we have
considered an exclusion statistics interpolating among FD, MB
and BE distributions.


\newpage
\begin{references}
\bibitem{Hal}
F.D.M.Haldane, Phys. Rev. Lett. {\bf 66}, 1529 (1991).
\bibitem{Wu}
Y.S. Wu, Phys. Rev. Lett. {\bf 73}, 922 (1994).
\bibitem{Lau}
R. B. Laughlin, Phys. Rev. Lett. {\bf 50}, 1395 (1983)
and in {\it The Quantum Hall Effect}, edited by R. Prange
and S. Girvin (Springer-Verlag, Heidelberg, 1989).
\bibitem{Fra}
{\it Fractional Statistics and Anyon Superconductivity}, edited
by F. Wilczek (World Scientific, Singapore, 1990).
\bibitem{Com}
{\it Common Trends in Condensed Matter and High Energy
Physics}, edited by L. Alvarez-Gaum\'e, A. Devoto,
S. Fubini, C. Trugenberger (North-Holland, Amsterdam, 1993).
\bibitem{Murt1}
M.V.N. Murthy and R. Shankar, Phys. Rev. Lett. {\bf 72}, 3629
(1994).
\bibitem{Murt2}
M.V.N. Murthy and R. Shankar, Phys. Rev. Lett. {\bf 73}, 3331
(1994).
\bibitem{Sen}
D. Sen and R.K. Bhaduri, Phys. Rev. Lett. {\bf 74}, 3912 (1995).
\bibitem{Isak}
S.B. Isakov, Phys. Rev. Lett. {\bf 73}, 2150 (1994).
\bibitem{Raja}
A.K. Rajagopal, Phys. Rev. Lett. {\bf 74}, 1048 (1995).
\bibitem{Nayak}
C. Nayak and F. Wilczek, Phys. Rev. Lett. {\bf 73}, 2740 (1994).
\bibitem{Ilin}
K.N. Ilinski and Gunn, preprint HEP-TH-9503233.
\bibitem{Achar}
R. Acharya and P. Narayana Swamy, J. Phys. A: Math. Gen.
{\bf 27}, 7247 (1994).
\bibitem{Poly}
A.P. Polychronakos, preprint UUITP-03/95, HEP-TH-9503077.
\bibitem{Gentile}
G. Gentile jr, Nuovo Cimento {\bf 17}, 493 (1940).
\bibitem{Ka}
G. Kaniadakis and P.Quarati, Phys. Rev. E {\bf 49}, 5103 (1994).
\bibitem{Landau}
L. D. Landau, E. M. Lifshitz, {\it Statistical Physics}, ( Pergamon
Press, London, 1959 ).

\end{references}
\end{document}